\documentclass[prb,twocolumn,superscriptaddress,showpacs,amsmath,amssymb]{revtex4}
\usepackage{amssymb}
\usepackage{txfonts}
\usepackage{bbm}
\usepackage{graphicx,graphics,color,epsfig}
\usepackage{appendix}
\usepackage{epstopdf}
\usepackage{appendix}

\begin{document}

\title{Insulator-metal transition and quasi-flat-band of Shastry-Sutherland lattice}

\author{Huai-Xiang Huang}
\email{hxhuang@shu.edu.cn}
\affiliation{Department of Physics, Shanghai University, Shanghai 200444, China}

\author{Jing Chen}
\affiliation{Department of Physics, Shanghai University, Shanghai 200444, China}

\author{Wei Ren}
\affiliation{International Centre for Quantum and Molecular Structures, Materials Genome Institute, Shanghai University, Shanghai 200444, China}

\author{Yi Gao}
\affiliation{Department of Physics and Institute of Theoretical Physics, Nanjing Normal University, Nanjing, Jiangsu 210023, China}

\author{Wei Li}
\email{w_li@fudan.edu.cn}
\affiliation{State Key Laboratory of Surface Physics and Department of Physics, Fudan University, Shanghai 200433, China}
\affiliation{Collaborative Innovation Center of Advanced Microstructures, Nanjing University, Jiangsu 210093, China}

\author{Yan Chen}
\affiliation{State Key Laboratory of Surface Physics and Department of Physics, Fudan University, Shanghai 200433, China}
\affiliation{Collaborative Innovation Center of Advanced Microstructures, Nanjing University, Jiangsu 210093, China}

\date{\today}

\begin{abstract}
Insulator-metal transition is investigated self-consistently on the frustrated Shastry-Sutherland lattice in the framework of Slave-Boson mean-field theory. Due to the presence of quasi-flat band structure characteristic, the system displays a spin-density-wave (SDW) insulating phase at the weak doping levels, which is robust against frustration, and it will be transited into an SDW metallic phase at high doping levels. As further increasing the doping, the temperature or the frustration on the diagonal linking bonds, the magnetic order $m$ will be monotonically suppressed, resulting in the appearance of a paramagnetic metallic phase. Although the Fermi surface of the SDW metallic phase may be immersed by temperature, the number of mobile charges is robust against temperature at weak doping levels.
\end{abstract}

 \maketitle

\section{Introduction}
Geometrically frustrated lattices~\cite{1,2,3,4,hxh,5,14prb,15prb} have exotic quantum states and rich phase diagrams since the antiparallel alignment of adjacent spins cannot be fully satisfied due to the energy competing. Among of them the Shastry-Sutherland lattice (SSL) is one of the simplest systems, which alternates diagonal links located on a square lattice. Compounds with topology equivalent to SSL, such as $\mathrm{SrCu_2(BO_3)_2}$, $\mathrm{Yb_2Pt_2Pb}$,~\cite{akoga,whz1,16,18,19,20,mk1,mk2,su} have been synthesized and provided an excellent platform to study the effects of frustration on the correlated electron systems.

Various analytical methods and numerical techniques have been employed to study the SSL.~\cite{dca,chc,ala,dav,sel,sha,haidi} From the viewpoint of localized two dimensional Heisenberg Hamiltonian, the original paper gave an exactly analytical solvable ground state~\cite{13}, which is the production of valence-bond dimer singlets on the disjointed diagonal links when the ratio of exchange coupling on diagonal bonds and that on $x$-axes $\alpha=J_{diag}/J_{x}>1.5$.~\cite{akoga,whz1} While for small $\alpha$, the N\'{e}el states will become the ground state since a SSL tends to degenerate to square lattice. Between the dimer-singlet state and the N\'{e}el state, many intermediate phases~\cite{mal,wzh2,pco,hxh2,zhen} and the magnetization plateaus~\cite{ssmag,ssmag1,ssmag2,Trinh} have also been reported.

The lately realized frustrated Lieb lattice~\cite{fla1,fla2,fla3,fla20,zhi} and the paradigmatic frustrated Kagom\'{e} lattice
have unpredicted properties owing to the flat-band structure.~\cite{fla18,ssmag,fla1} Moreover, the frustrated graphene sheets lead the system displaying a Mott-like insulator due to the strong electron correlation.~\cite{yuan} SSL can be constructed as a quasi-flat-banded, frustrated and correlated electron system based on $t_1$-$t_2$-$J_1$-$J_2$ model and it motivates us to study the intriguing metal-insulator transition systematically.

Considering the itinerant electron behaviors, the $t$-$J$ model has been used to study the possible superconducting phase on SSL,~\cite{hxh2,chunghouchung} and the Hubbard model has been used to clarify the metal-insulator transition at around the half-filling.~\cite{haidi} However, insulator-metal phase transition on SSL at finite doping and the finite temperature is still lacking investigation theoretically as far as we have known.

In the present paper, we mainly focus on the study of intriguing insulator-metal phase transition on SSL by using a $t_1$-$t_2$-$J_1$-$J_2$ model in the framework of Slave-Boson (SB) approach.~\cite{ale,cts,kot,yuanqsh} Once the mean-field order parameters are self-consistently calculated, the band structure and Fermi surface topology can be evaluated straightforwardly. Based on the linear-response theory, the temperature dependent Drude weight proportionally to the electrical conductivity is also investigated. The effect of the finite third nearest-neighbor (n.n.) hopping term $t_3$ on the flat band structure will also be addressed in the appendix.

Besides the unfrustrated square lattice, lots of unpredicted properties will arise from the disjointed diagonal frustrated parameters $t_2$ and $J_2$ on the SSL. Strong electron-electron interaction will separate the energy bands into two subspaces at weak dopings, and the system displays an insulating state with SDW order (SDW-Ins) and absence of Fermi surface, which is robust against $t_2$, $J_2$ with tiny Drude weight. At high doping levels, the system enters into a metallic phase, which is separated as SDW metallic (SDW-M) phase and paramagnetic metallic (PM-M) phase depending on whether the staggered magnetic order is finite.

The rest of this paper is organized as follows. In Sec.~\ref{model}, we will introduce the theoretical model and the corresponding analytical formalism. The detailed calculated phase diagram is shown in Sec.~\ref{phasediagram}. In Sec.~\ref{temperature_effects}, we will turn to discuss the temperature dependent effects on the calculated physical quantities. At last, a summary is given in Sec.~\ref{summary}.

\section{Model Hamiltonian and formula}\label{model}

\begin{figure}
\centering
      \includegraphics[width=8.5cm]{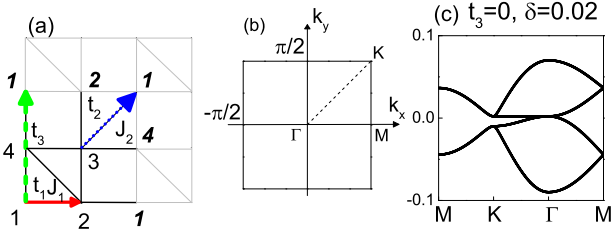}
\caption{(Color Online) (a) Schematic illustration of SSL, $i=1..4$ denoting the four different sublattices. Order parameters on the dark links belong to the same unit cell. Note that $1\rightarrow2$ and $2\rightarrow\textbf{\emph{1}}$ are two different links in one unit cell. $t_1$ , $J_1$ are on the n.n. links denoted by the red solid line, $t_2$, $J_2$ are on the diagonal links denoted by the blue dotted line, and $t_3$ on the third n.n. is denoted by the green dashed line. (b) The first Brillouin zone of SSL. $\Gamma$, M, K are the three high symmetry $\mathbf{k}$-points. (c) The energy band dispersion of the tight-binding model along the high symmetric line of M-K-$\Gamma$-M for the parameters of $t_1=\pm 1$, $t_2=0.3$, $t_3=0$ and $\delta=0.02$.}\label{fig1}
\end{figure}

We consider the $t_1$-$t_2$-$t_3$-$J_1$-$J_2$ phenomenological model to systematically study the frustrated SSL by using SB mean-field theory, where $t_3$ is set to zero without explicitly specified.
The crystal structure of SSL is illustrated in Fig.~\ref{fig1}(a). Within a unit cell, there are four inequivalent sites labeled by $1-4$, the hopping integral $t_1$ and spin exchange coupling $J_1$ are on the n.n. links, and $t_2$ and $J_2$ are on the diagonal links. Generally, the $t$-$J$ model can be derived analytically from the Hubbard model in the strong coupling limit with $J=4t^2/U$, where $U$ is the on-site Coulomb interaction in the Hubbard model. The generalized $t$-$J$ model with independent parameters of $t$ and $J$ can be used to describe the more complex systems, like SSL from the phenomenological viewpoint.
In real space the Hamiltonian reads
\begin{eqnarray}\label{Hamreal}
&H&=H_t+H_J-\mu_0\sum_i n_i=-t_1\sum_{\langle ij \rangle,\sigma}(\hat{c}^{\dag}_{i\sigma }\hat{c}_{j\sigma }+h.c.)\nonumber\\
&&-t_2\sum_{\langle ij \rangle_2,\sigma}(\hat{c}^{\dag}_{i\sigma }\hat{c}_{j\sigma }+h.c.)-t_3\sum_{\langle ij \rangle_3,\sigma}(\hat{c}^{\dag}_{i\sigma }\hat{c}_{j\sigma }+h.c.)\nonumber\\
&+&J_{1}\sum_{ \langle ij \rangle}  (\hat{S}_i \cdot \hat{S}_j-\frac{1}{4}\hat{n}_i\hat{n}_j)+J_{2}\sum_{ \langle ij \rangle_2 }  (\hat{S}_i \cdot \hat{S}_j-\frac{1}{4}\hat{n}_i\hat{n}_j)\nonumber\\
&-&\mu_0\sum_i n_i,
\end{eqnarray}
where $\sigma$ denotes spin,  $\langle \rangle$, $\langle \rangle_2$, $\langle \rangle_3$ represent the n.n., second n.n. and third n.n. sites, respectively, as shown in Fig.1(a), with
$\hat{S}_i=\frac{1}{2}(\hat{c}^{\dag}_{i\alpha}\mathbf{\sigma}_{\alpha\beta}\hat{c}_{i\beta})$ denoting the spin operator. For a strong coupling limit and hole-doped case, $\hat{c}_{j\sigma }=c_{j\sigma }(1-\mathfrak{n}_{j\bar{\sigma}})$ means annihilating an electron on a single occupied site with $(1-\mathfrak{n}_{j\bar{\sigma}})\simeq1$, and $\hat{c}^{\dag}_{i\sigma }=(1-\mathfrak{n}_{i\bar{\sigma}})c^{\dag}_{i\sigma }$ represents creating an electron on an empty site with $(1-\mathfrak{n}_{i\bar{\sigma}})\simeq \delta$, $\mathfrak{n}$ representing the occupation number of real electrons and $\delta$ describing the hole-doped concentration. SB~\cite{ale,cts,kot,yuanqsh} approach allows a physical description of the electron correlated effects, by writing $\hat{c}_{i\sigma}=b^{\dag}_if_{i\sigma}$ with $b_i$ Boson holon operator and $f_{i\sigma}$ the Fermionic spinon operator.
Without double occupancy constraint, $b_i^{\dag} b_i+\Sigma_{\sigma} f_{i\sigma}^{\dag}f_{i\sigma}=1$ has to be satisfied and is imposed on the Hamiltonian through a Lagrange multiplier, thus $\mu_0$ in Eq. (\ref{Hamreal}) changes to $\mu$ hereafter, which is the chemical potential and controls the electronic concentrations $n_i=\sum_{\sigma}n_{i\sigma}=\sum_{\sigma}f^{\dag}_{i\sigma}f_{i\sigma}$. After defining the staggered magnetic order $m=(-1)^{i}\langle n_{i\uparrow}-n_{i\downarrow}\rangle/2$, we obtain the relation $\langle n_{i\sigma}\rangle=(1-\delta)/2+\sigma(-1)^i m$.
In the static SB approximation, the boson condensation is assumed $\langle b_i^{\dag} b_i\rangle=\delta$ since the bosonic fluctuations are suppressed.

Introducing the mean-field bond order $\chi^{\sigma}_{ij}=\langle f^{\
\dag}_{i\sigma}f_{j\sigma}\rangle$, spin coupling interaction in mean-field level reads
\begin{eqnarray}\label{Hamj}
H_J&=&-\Sigma_{\langle ij \rangle,\langle ij \rangle_2,\sigma} \frac{J_{ij}}{2}[c^{\dag}_{i\sigma}c^{\dag}_{j\bar{\sigma}}c_{i\bar{\sigma}}c_{j\sigma}-c^{\dag}_{i\sigma}c^{\dag}_{j\bar{\sigma}}c_{i\sigma}c_{j\bar{\sigma}}]\nonumber\\
&=&-\Sigma_{\langle ij \rangle,\langle ij \rangle_2, \sigma} \frac{J_{ij}}{2}[\langle n_{i\sigma}\rangle f^{\dag}_{j\bar{\sigma}}f_{j\bar{\sigma}} +\langle n_{j\sigma}\rangle f^{\dag}_{i\bar{\sigma}}f_{i\bar{\sigma}}\nonumber\\
&+&\chi^{\sigma}_{ij}f^{\dag}_{j\bar{\sigma}}f_{i\bar{\sigma}}+\chi^{\sigma}_{ji}f^{\dag}_{i\bar{\sigma}}f_{j\bar{\sigma}}],
\end{eqnarray}
where $J_{ij}$ equals $J_1$ on the n.n. links and $J_2$ on diagonal links, otherwise it is zero, shown in Fig.~\ref{fig1}(a). Since we are interested in the insulator-metal transition, the superconducting order is ignored for simplify.~\cite{pwl}
In the momentum $\mathbf{k}$ space, the Hamiltonian is expressed as $\psi^{\dag} [H_k-\mu]\psi$ after dropping a constant term, where $\psi^{\dag}=[f^{\dag}_{1\uparrow}(k)...f^{\dag}_{4\uparrow}(k),f^{\dag}_{1\downarrow}(k)...f^{\dag}_{4\downarrow}(k)]$, and
\begin{eqnarray}\label{1}
 H_k&=&\left(
 \begin{array}{cc}
 H_{t,k}+ H^{\uparrow}_{J,k} & 0 \\
 0 & H_{t,k}+H^{\downarrow}_{J,k} \\
 \end{array}
 \right),\\
 H_{t,k}&=&\delta \left(
  \begin{array}{cccc}
     t_3a_3    &-2t_1\cos{k_x}     & -t_2 e^{-\mathrm{i}(k_x+k_y)}  & -2t_1\cos{k_y} \\
    -2t_1\cos{k_x}      & t_3a_3    & -2t_1\cos{k_y}  & -t_2 e^{-\mathrm{i}(k_x-k_y)} \\
    -t_2 e^{\mathrm{i}(k_x+k_y)} & -2t_1\cos{k_y} & t_3a_3 & -2t_1 \cos{k_x} \\
    -2t_1\cos{k_y}   & -t_2 e^{\mathrm{i}(k_x-k_y)} & -2t_1\cos{k_x} & t_3a_3 \\
  \end{array}
\right), \nonumber
\end{eqnarray}
with $a_3=-2(\cos{2k_x}+\cos{2k_y})$, and $\mathbf{k}$ is restricted in the first Brillouin zone as shown in Fig.~\ref{fig1}(b). In the deduction $\langle b^{\dag}_ib_j \rangle$ is approximated as $\delta$, and $b_ib_jb^{\dag}_ib^{\dag}_j$ is replaced by unity. In the following $|t_1|$ and the distance between the n.n. site are set as  energy unit and  length unit, respectively.
It is easy to prove that $\pm t_1$ will give the same energy band dispersions for the tight-binding model $H_{t,k}$.
For the special range $k_x=k_y$, one of the band dispersion of $H_{t,k}$ has a simply analytical expression as
\begin{eqnarray}\label{2}
\epsilon=\delta( t_3a_3+t_2),
\end{eqnarray}
which is a constant for $t_3=0$, indicating a quasi-flat band represented by the flat segment from K to $\Gamma$ shown in Fig.~\ref{fig1}(c).

Interaction Hamiltonian $ H^{\sigma=\uparrow,\downarrow}_{J,k}$ depends on spin indexes, for example $H^{\sigma}_{J,k}(1,2)=-0.5J_1  (\chi^{\bar{\sigma}}_{21}e^{\mathrm{i}k_x}+\chi^{\bar{\sigma}}_{2\textbf{\emph{1}}}e^{-\mathrm{i}k_x} )$, $H^{\sigma}_J(2,4)=-0.5J_2 \chi^{\bar{\sigma}}_{42}e^{-\mathrm{i}(k_x-k_y)}$,
where the $\chi^{\sigma}_{21}$ and $\chi^{\sigma}_{2\textbf{\emph{1}}}$ are on different links shown in the Fig.~\ref{fig1}(a), and the diagonal element $H^{\sigma}_{J,k}(1,1)$ reads
$-J_1(1-\delta-\sigma2m)-0.25J_2(1-\delta+\sigma2m)$. In the Hamiltonian $m$, $\mu$ as well as the $20$ mean-field $\chi^{\sigma}_{ij}$ are self-consistently calculated.

In the presence of a slowly varying vector potential $\textbf{A}$
along $x$-direction, the associated Peierls phase is $ c^{\dag}_{i\sigma}c_{j\sigma}\exp{\mathrm{i}\frac{e}{\hbar
c}\int^{r_i}_{r_j}\textbf{A}(\textbf{r},t)\cdot \mathrm{d} \textbf{r}}$, and the charge current density can be decomposed into diamagnetic and paramagnetic part $J_x(r_i)=eJ^{p}_x(r_i)+e^2K_x(r_i)\textbf{A}_x(r_i,t)$, where
\begin{eqnarray}\label{curr}
K_x(r_i)&=&-\delta\sum_{\sigma j}t_{ij}(f^{\dag}_{i\sigma}f_{j,\sigma}+H.c.),\\
J^P_x(r_i)&=&-\mathrm{i}\delta\sum_{\sigma j}t_{ij}(f^{\dag}_{i\sigma}f_{j,\sigma}-H.c.),
\end{eqnarray}
with $j=i+\vec{x}, i+\vec{x}\pm \vec{y}$. Applying the linear response theory, the Drude weight $\frac{D}{\pi e^2}=\frac{1}{N}\Pi_{xx}(\textbf{q}=0,\omega\rightarrow 0)-\langle K_x \rangle_0$, a measurement of the ratio of density of mobile chargers to their mass, can be evaluated straightforwardly.~\cite{1d,2d,3d,4d,5d} The current-current correlation reads $\Pi_{xx}(\textbf{q},\tau)=-\langle T_{\tau} J^P_x(\textbf{q},\tau)J^P_x(-\textbf{q},0)\rangle_0$ with $J^P_x(\textbf{q},\tau)=e^{\tau H}J^P_x(\textbf{q}) e^{-\tau H}$, $\mathrm{T}_{\tau}$ time ordering operator, $\tau$ imaginary time, and $J^P_x(\textbf{q})=\sum_i e^{-\mathrm{i}\textbf{q}\cdot r_i}J^P_x(r_i)$. Here it should be noted that for an insulating phase $D$ is close to zero, whereas $D$ is finite for a metallic phase.

The diamagnetic current $K_x$ is easy to derive since it needs to be calculated to zeroth order of $\textbf{A}$. The current correlated function is expressed as  $\Pi_{xx}(q,\tau)=\frac{1}{\beta}\sum_{n}\Pi_{xx}(q,\mathrm{i}\omega_n)e^{-\mathrm{i}\omega_n\tau}$
in the Matsubara formalism. It can be expressed as
\begin{eqnarray}\label{para}
\Pi_{xx}(\textbf{q},\mathrm{i}\omega)
            \!\!=\!\!\!\!\!\sum_{{\bf{k}}m_1m_2} \!\! \frac{ Y_{m_1m_2}(k)Y_{m_2m_1}(k)(\mathbb{F}(E_{{\bf{k}},m_1})-\mathbb{F}(E_{{\bf{k}},m_2}) ) }{\mathrm{i}\omega+(E_{{\bf{k}},m_1}-E_{{\bf{k}},m_2})},
\end{eqnarray}
by using the equation of motion, where $\mathbb{F}$ is the Fermi distribution function, $Y_{m_1m_2}(k)$ is a lengthy straightforward function of transformation matrix $\mathbb{T}$, obtained in the process of diagonalizing $H_k$. Through analytic continuation of $\Pi_{xx}(q,\mathrm{i}\omega)|_{i\omega=\omega+i\eta}$, $\Pi_{xx}(\textbf{q},\omega)$ appearing in the expression of $D$ is obtained.

Throughout the this paper, we use the parameters of $t_1=\pm1$ and $J_1=0.3$, by following the previous studies on the square lattice,~\cite{yuanqsh,tto,tkl,pwl} and these choices do not affect on the physical discussions. The number of unit cell in the self-consistent calculation is $N=128\times 128$ with the accuracy less than $10^{-4}$, while in the calculations of density of state (DOS), Drude weight $D$ and the band structure are used $N=640\times 640$.
Since there are $16$ n.n. bond orders, which can be divided into two groups with $ \chi^{\uparrow}_{4\textbf{\emph{1}}}=\chi^{\uparrow}_{3\textbf{\emph{2}}}=\chi^{\uparrow}_{3\textbf{\emph{4}}}=\chi^{\uparrow}_{2\textbf{\emph{1}}}=
\chi^{\downarrow}_{43}=\chi^{\downarrow}_{23}=\chi^{\downarrow}_{12}=\chi^{\downarrow}_{14}$ and $ \chi^{\downarrow}_{4\textbf{\emph{1}}}=\chi^{\downarrow}_{3\textbf{\emph{2}}}=\chi^{\downarrow}_{3\textbf{\emph{4}}}=\chi^{\downarrow}_{2\textbf{\emph{1}}}=
\chi^{\uparrow}_{43}=\chi^{\uparrow}_{23}=\chi^{\uparrow}_{12}=\chi^{\uparrow}_{14}$ having a negligible difference, we use the average of them as parameter $\chi_{x,y}=\frac{1}{16}\sum_{\sigma i}\chi^{\sigma}_{ij=i+\vec{x},\vec{y}}, i=1..4$ for simplicity. Similarly, the average bond order on the diagonal links is denoted as $\chi_d=\frac{1}{4}\sum_{ \sigma}\chi^{\sigma}_{2,4}+\chi^{\sigma}_{3,\textbf{\emph{1}}}$.
Although we use $\chi_{x,y}$ and $\chi_d$ as parameters in the plots, in all calculations the exact $\chi^{\sigma}_{ij}$ are used.

\section{phase diagram at $T=0.001$}\label{phasediagram}

\begin{figure}
      \includegraphics[width=8.5cm]{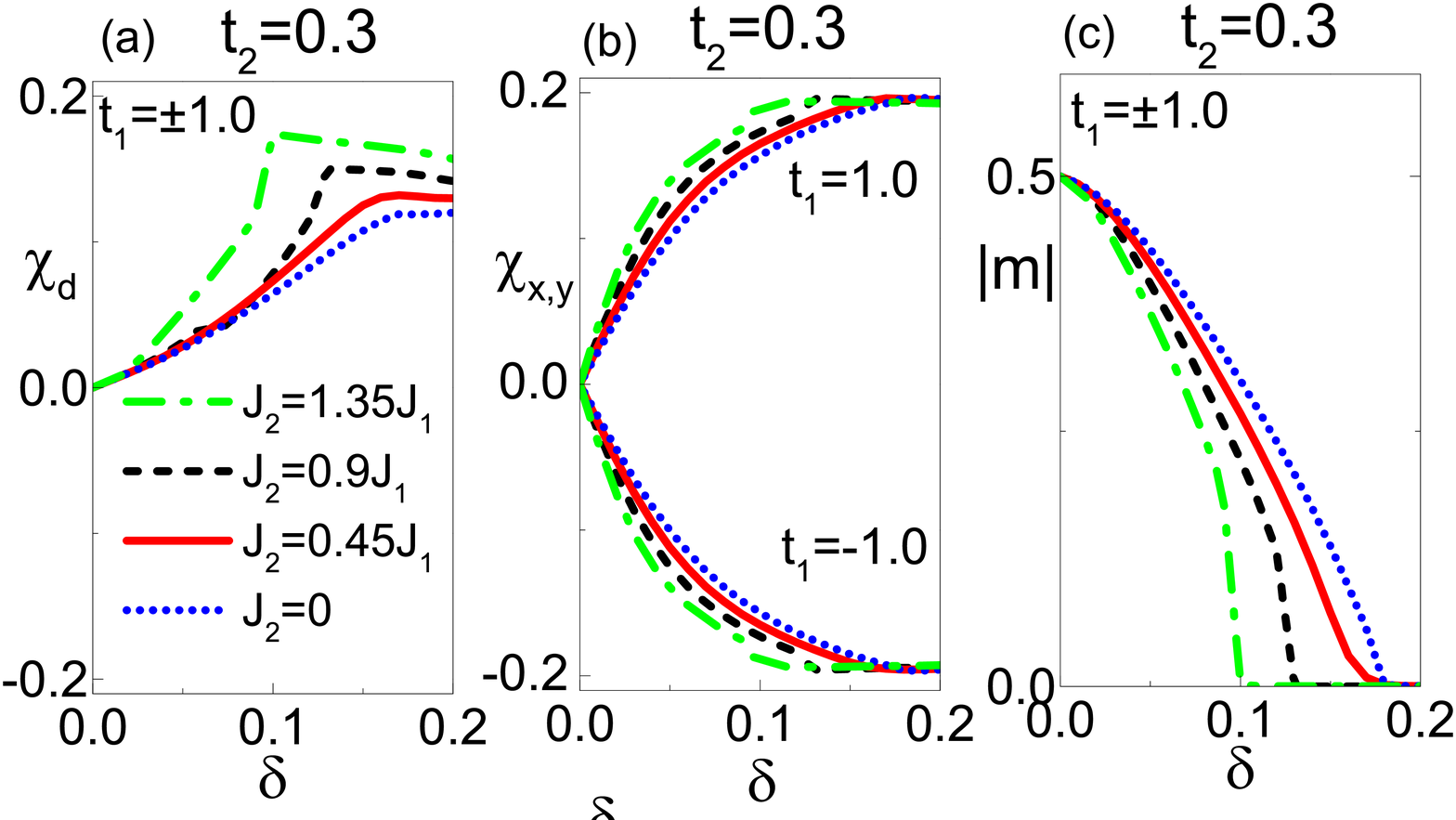}
\caption{(Color Online) Doping dependent order parameters for fixed $t_2=0.3$ and four different $J_2$. (a) $\chi_{d}$ as function of $\delta$ for $t_1=\pm1.0$. (b) $\chi_{x,y}$ as function of $\delta$, the upper lines are for $t_1=1$ and the lower lines are for $t_1=-1.0$. (c) $|m|$ as function of $\delta$ for $t_1=\pm1.0$. }\label{fig4}
\end{figure}

\begin{figure}
\centering
      \includegraphics[width=4cm]{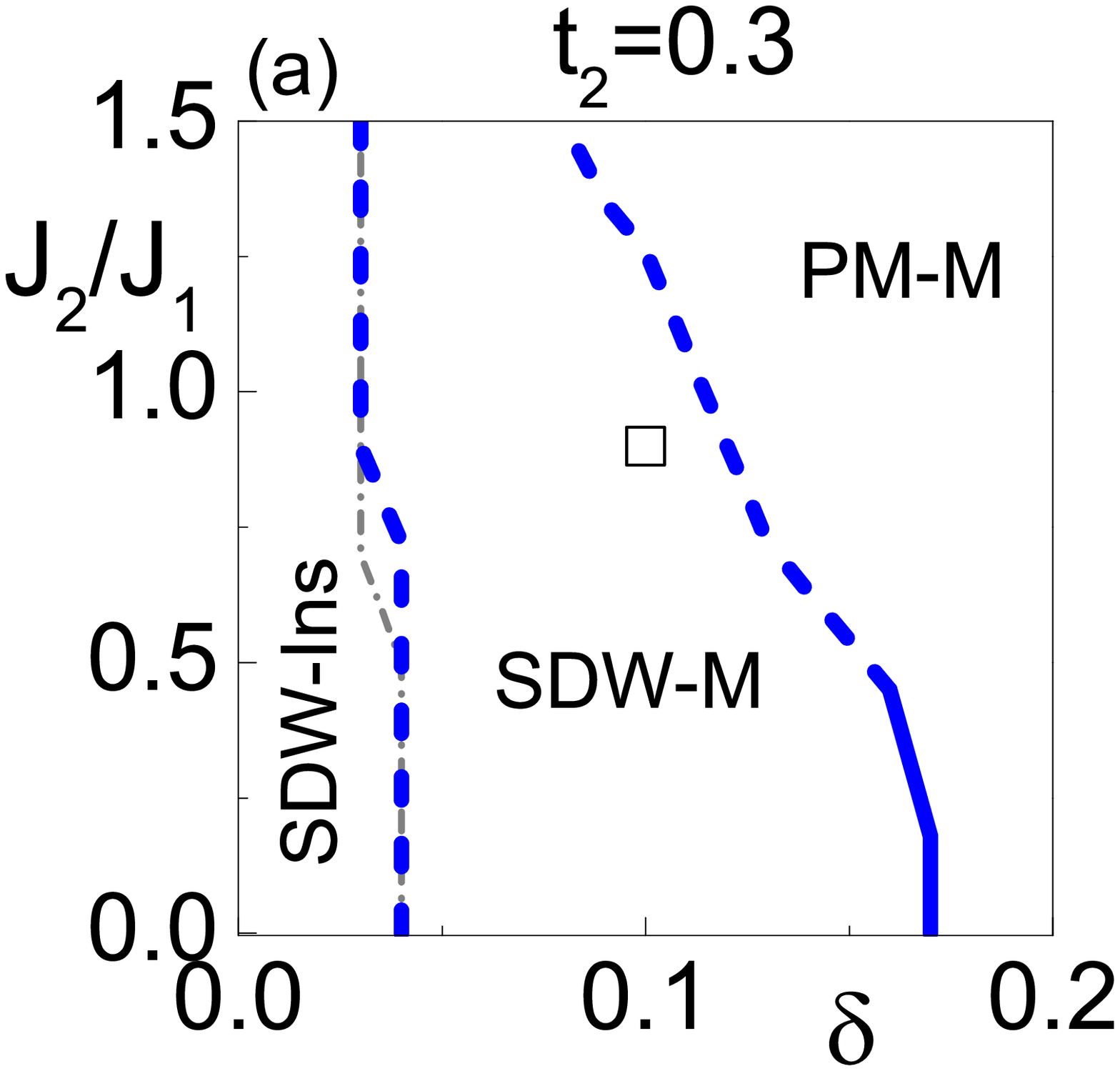}
      \includegraphics[width=4cm]{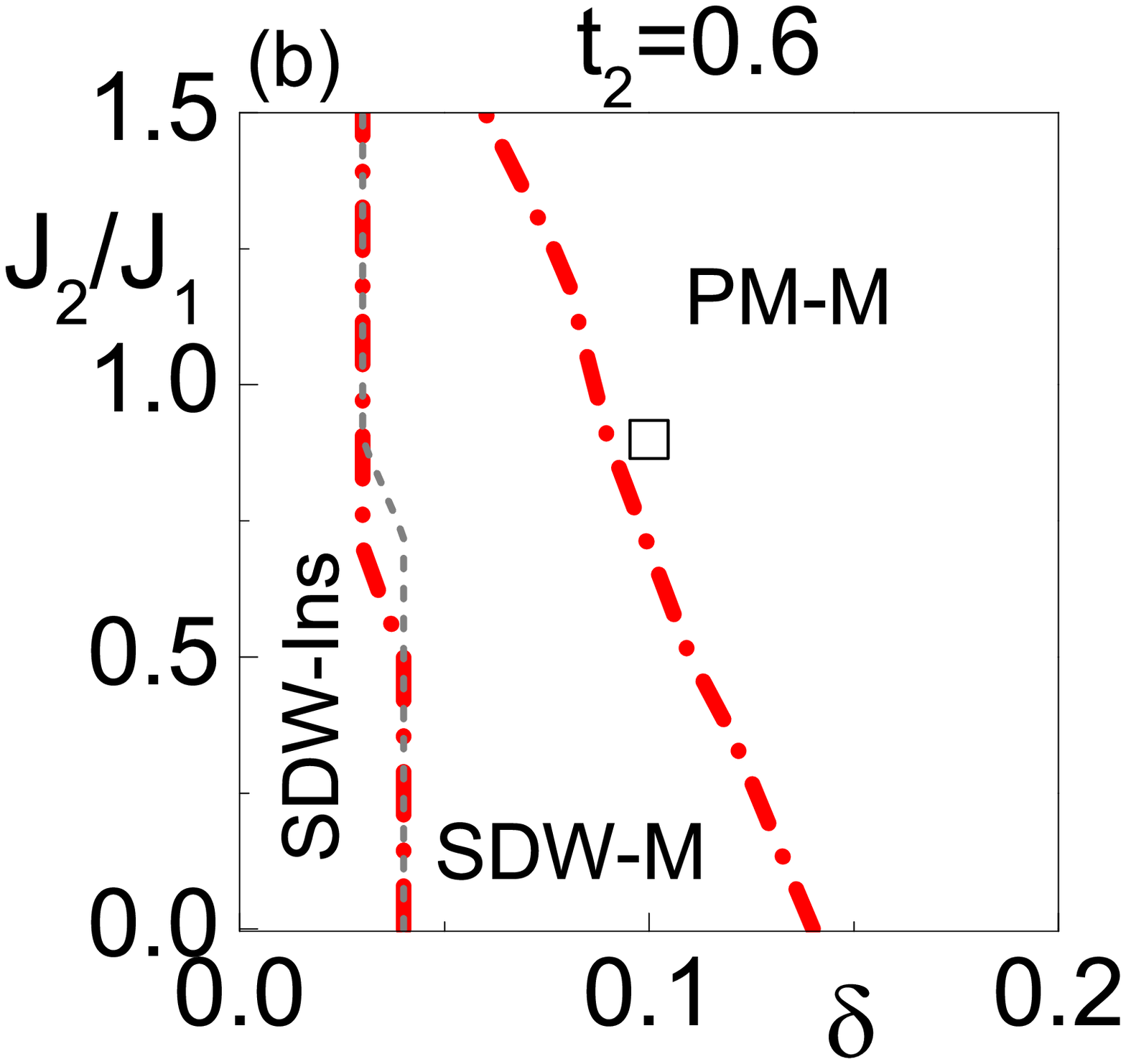}
      \includegraphics[width=4cm]{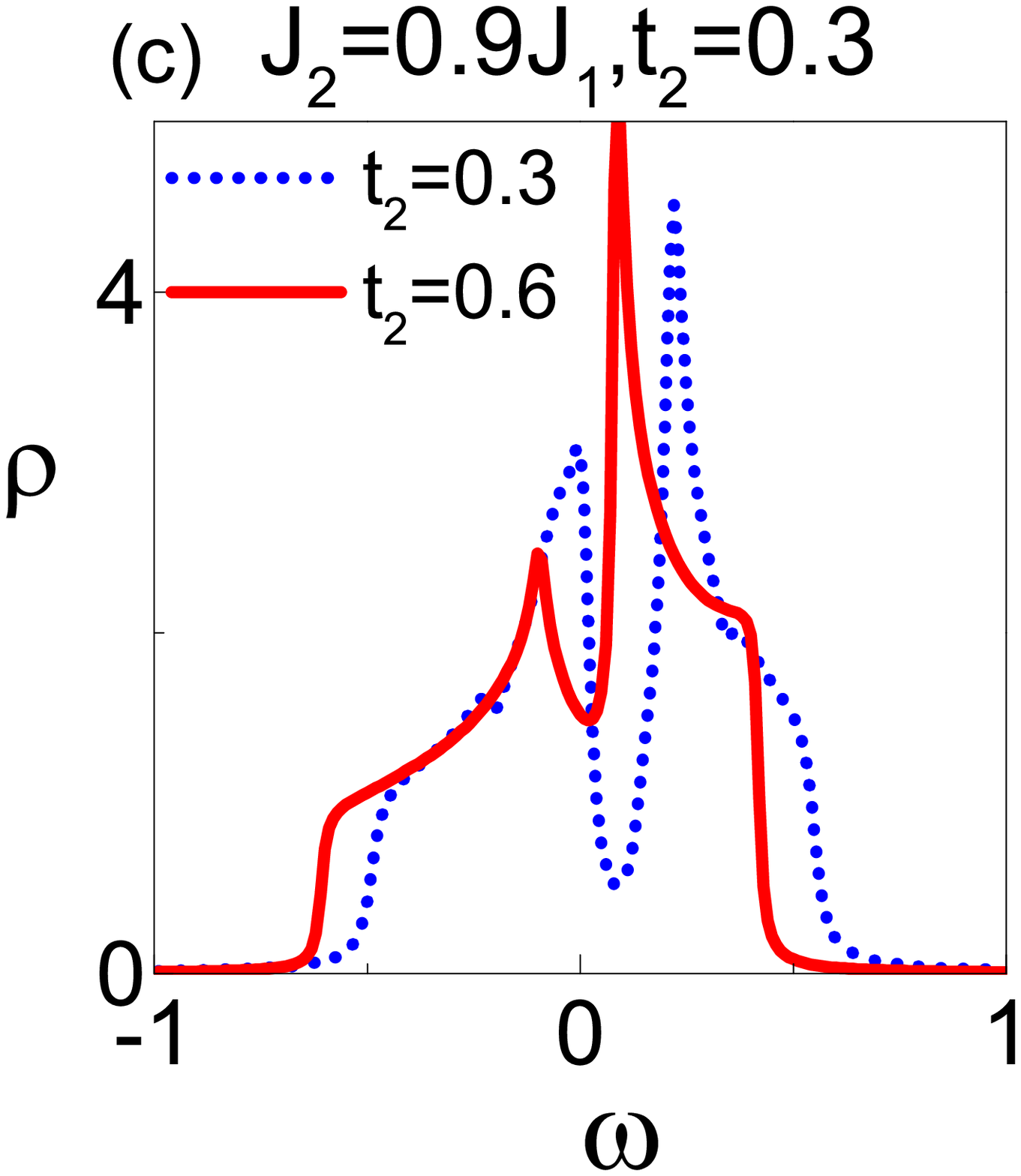}
      \includegraphics[width=4cm]{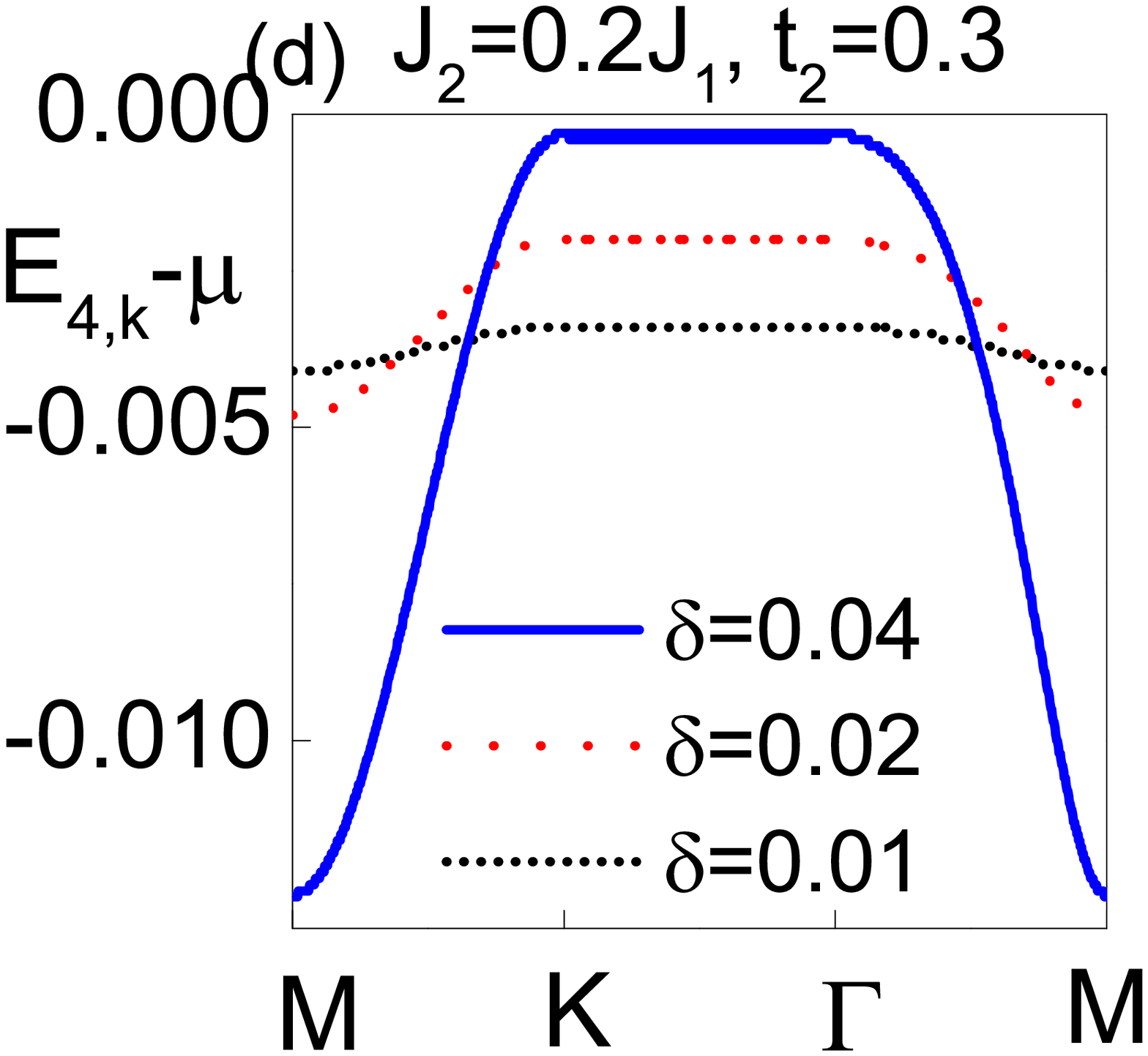}
\caption{(Color Online) (a) Phase diagram of $t_1$-$t_2$-$J_1$-$J_2$ model on the $J_2/J_1$-$\delta$ plane for $t_2=0.3$. The blue solid line corresponds to the first order phase transition. (b) Phase diagram for $t_2=0.6$. The grey dashed line in $t_2=0.3(0.6)$ is the boundary of SDW-Ins phase for $t_2=0.6 (0.3)$.
(c) DOS of the point $J_2=0.9J_1,\delta=0.1$ for $t_2=0.3,0.6$, respectively.
(d) Energy dispersion of the quasi-flat band in the lower subset along the line of M-K-$\Gamma$-M for $ t_2=0.3$, $J_2=0.2J_1$. From down to up the curves corresponds to $\delta=0.01,0.02,0.04$, respectively.}\label{fig5}
\end{figure}

In this section we will discuss the phase diagram of $t_1$-$t_2$-$J_1$-$J_2$ model at $T=0.001$. With fixed $t_2=0.3$, Fig.~\ref{fig4} depicts the doping dependent mean-field order parameters for four different values of $J_2$.
It shows that the diagonal bond order $\chi_{d}$ and the magnetic order $|m|$ are equal for $t_1=\pm1$, whereas $\chi_{x,y}$ is equal in magnitude but has plus sign for $t_1=1$ and minus sign for $t_1=-1$.
The sign change of the mean-field $\chi_{x,y}$ along the short bond order under a hopping parameter switch $t_1\rightarrow-t_1$ shows that the band structure remains unchanged for $t_1=\pm 1$, therefore the sign of $t_1$ is an irrelevant quantity in our investigations.
The large $J_2$ prefers to enhance $|\chi|$ but suppress $|m|$, which is ascribed to the presence of competition between kinetic energy and magnetic order. From Fig.~\ref{fig4}(c) we can see that for $J_2=0$, $|m|$ versus $\delta$ is a smooth curve, while for  $J_2>0.45J_1$, $|m|$ drops to zero abruptly and corresponds to the occurrence of the first order phase transition. At zero doping all kinds of $\chi=0$, finite doping leads to finite $\chi$ ascribing to the characteristic of itinerate electrons.

\begin{figure}
\centering
       \includegraphics[width=7.2cm]{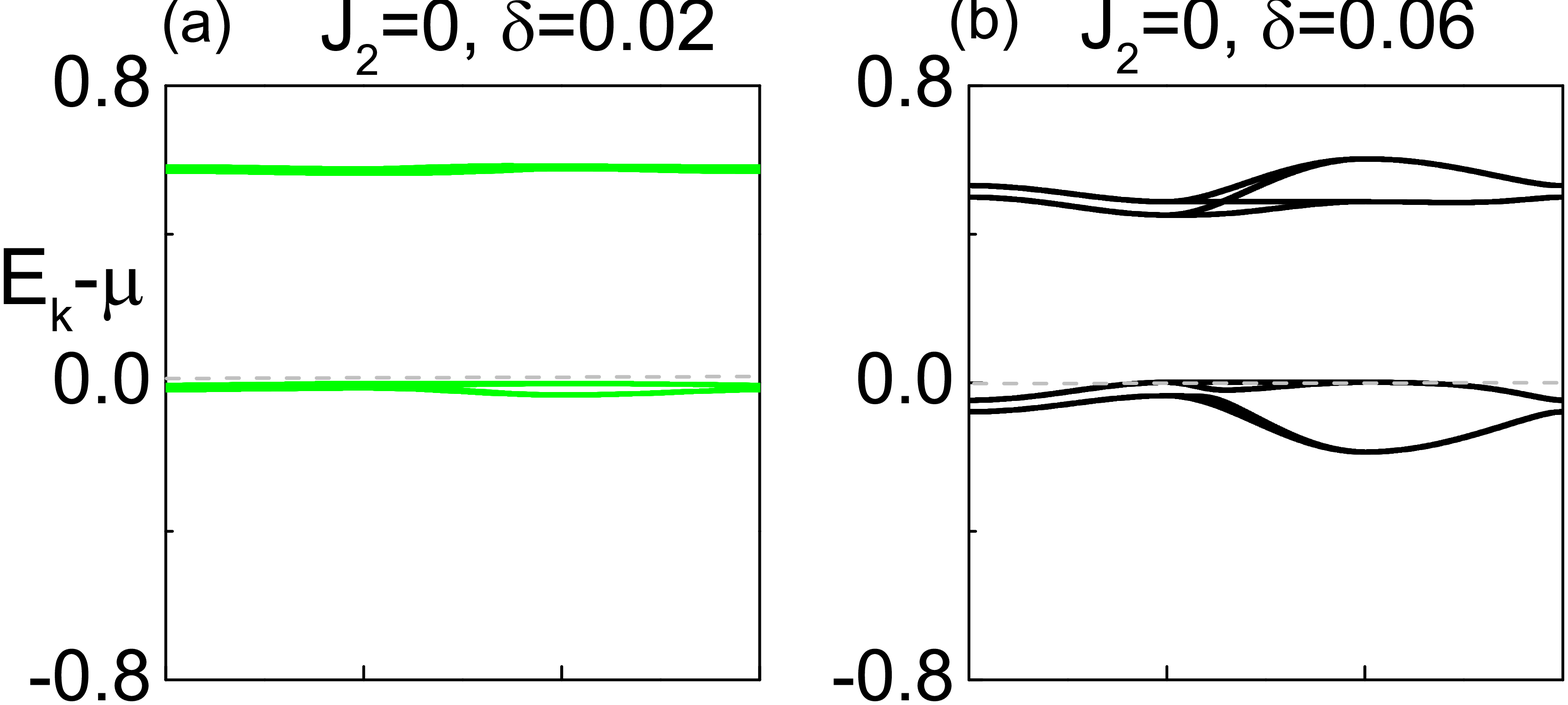}
       \includegraphics[width=7.2cm]{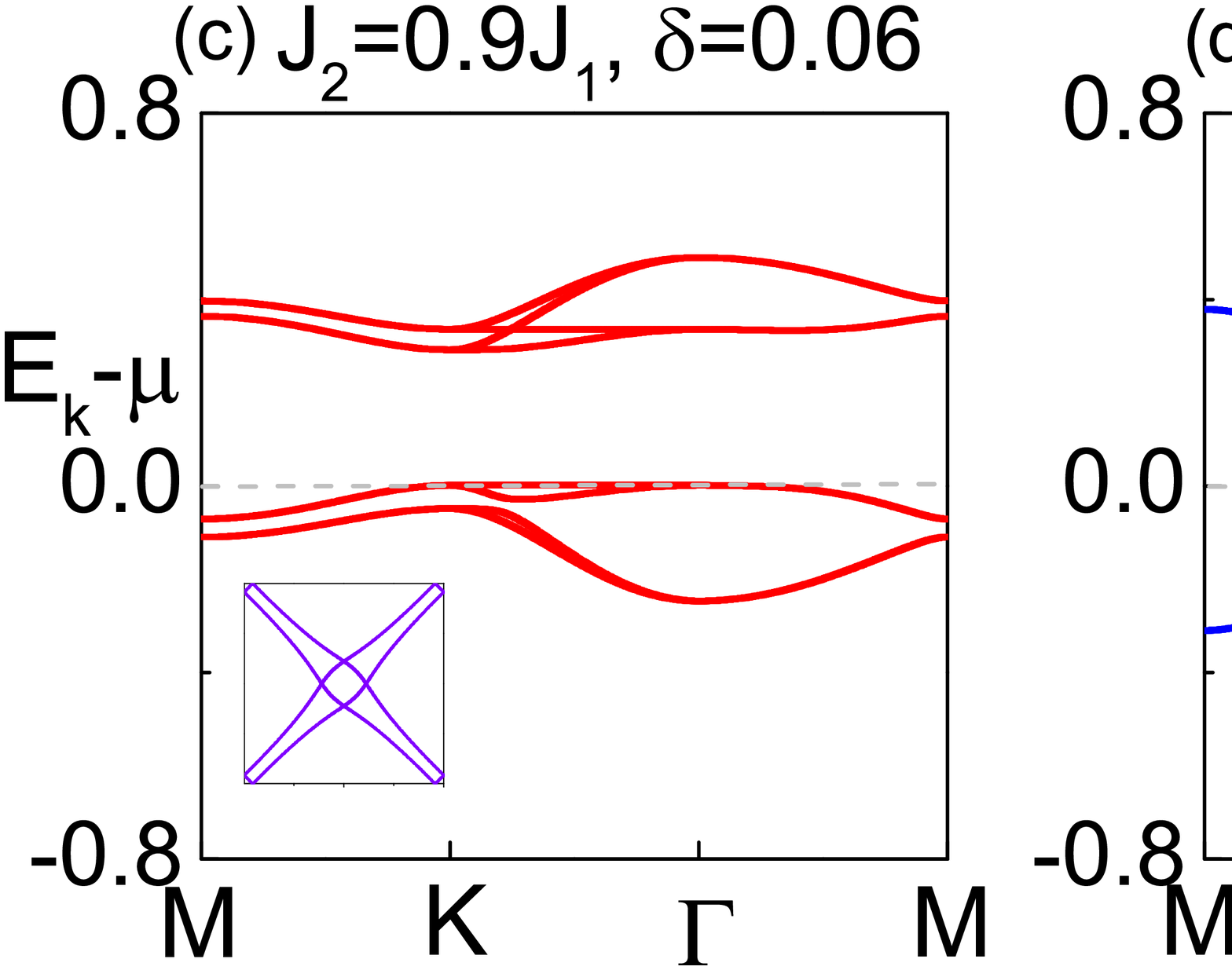}
\caption{(Color Online) For $t_2=0.3$, the energy band dispersions along the high symmetric line M-K-$\Gamma$-M for the four selected set of $J_2$ and doping. The grey dashed lines denote the Fermi level. The inset of (c) is the corresponding Fermi surface topology. }\label{fig6}
\end{figure}

Fig.~\ref{fig5}(a) and (b) show the phase diagram on the $J_2/J_1$-$\delta$ plane for fixed $t_2=0.3$ and $0.6$, respectively. The remarkable feature is that the area of SDW-Ins phase almost remains the same for different values of $t_2$, while the area of SDW-M phase shrinks as increasing the $t_2$. The point $J_2=0.9J_1,\delta=0.1$ denoted by the hollow square locates in the SDW-M phase for $t_2=0.3$, where the corresponding DOS $\rho$ is finite at $\omega=0$ with two pronounced peaks located at the edge of the spin gap [blue dotted line in Fig.~\ref{fig5}(c)], while it locates in PM-M phase for $t_2=0.6$, with a pronounced in-gap resonance peak arising [red solid line in Fig.~\ref{fig5}(c)], suggesting a highly DOS due to the vanishing $|m|$.

The strong electron-electron interaction separates the energy bands into two subspaces at small dopings. For $J_2=0.2J_1$ and $t_2=0.3$, Fig.~\ref{fig5}(d) shows the band dispersion along the line of high symmetric M-K-$\Gamma$-M for the quasi-flat band $E_{4,k}$,
which is the top band in the lower subspace. We find that in the SDW-Ins phase, a finite distance between the low-energy bands and the Fermi level exists, and $E_{4,k}$ is much closer to the Fermi level for higher doping. Numerical calculations show that on the flat segment electron concentration is $\langle n_p(k)\rangle=0.968, 0.881, 0.579$ for $\delta=0.01, 0.02, 0.04$, which equals to the Fermi distribution function.

\begin{figure}
      \includegraphics[width=8cm]{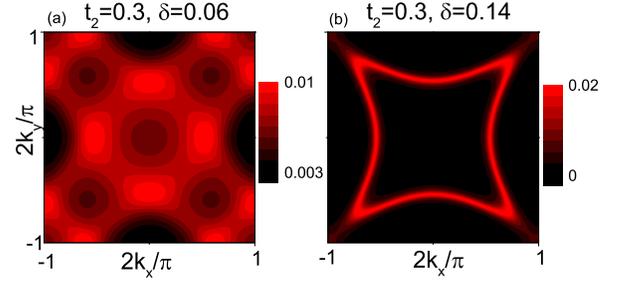}
\caption{(Color Online) The map of spectral weight of $\omega=0$ for $t_2=0.3$ and $J_2=0.9J_1$ in the $\mathbf{k}$ space with (a) corresponding to SDM-M phase at $\delta=0.06$ and (b) corresponding to PM-M phase at $\delta=0.14$.
}\label{ferm1}
\end{figure}

Taking $t_2=0.3$ as an example, we show the energy band structure in the whole energy range in Fig.~\ref{fig6}. For an undoped case, the system displays an insulating phase. For $J_2=0$ and $\delta=0.02$, the system is in SDW-Ins phase with band structure looking like two straight lines. At $\delta=0.06$ [see panel (b)], the system is in SDW-M phase, and the eight bands are well separated with the flat segment of $E_{4,k}$ slightly above the Fermi level. As shown in Fig.~\ref{fig6} (b) and (c), the larger frustration $J_2$ will
gradually suppress the height of the upper bands and diminish the SDW gap. The inset of Fig.~\ref{fig6} (c) shows the corresponding Fermi surface of the SDW-M phase. For $J_2=0.9J_1$ and $\delta=0.14$, the low and high energy bands will be touched, and the system will be driven into the PM-M phase with the eight bands degenerated into four bands. In all cases, the flat segment from M to $\Gamma$ remains the same features.

The distribution function of occupation numbers in the $\mathbf{k}$ space can be expressed as $n_k=\int d\omega A(k,\omega)f(\omega)$, where the spectral function
$A(k,\omega)=-\frac{\delta}{\pi}\mathrm{Im} \frac{1}{4}\sum_i\mathcal{G}_i(\textbf{k},\mathrm{i}\omega)|_{\mathrm{i}\omega=\omega+\mathrm{i}0^{+}}$ is the imaginary part of the single-particle Green's function multiplied by $-1/\pi$, giving the correlation of the electron creation and annihilation operations. For each sublattice $\mathcal{G}_i(\textbf{k},\mathrm{i}\omega)=\int^{\beta}_0 -T_{\tau}\langle f_{i k}(\tau)f^{\dag}_{i k}(0) \rangle e^{\mathrm{i}\omega\tau}$ is a function of  transformation matrix and $\mathcal{G}^0_n(\textbf{k},\mathrm{i}\omega)=1/(\mathrm{i}\omega-E_{n,k}))$.
For an interaction system single-particle Green's function or $A(k,\omega)$ carries information about the underlying potential and no longer concentrated at single energy like that in non-interaction systems. Experimentally, the angle-resolved photoemission spectroscopy (ARPES) experiment is often used to detect the single-particle information and the Fermi surface topologies. To qualitatively compare with the ARPES measurements, the spectral weight is evaluated, which is obtained by integrating the spectral function $A(k,\omega)$ times the Fermi distribution function over an energy interval of $[-0.01,0.01]$ around $\omega$. The corresponding Fermi surface topology is set $\omega=0$ and shown in Fig.~\ref{ferm1}. In the SDW-Ins phase, the map of spectral weight is invisible, which is a typical feature of insulating state. For $t_2=0.3$, $J_2=0.9J_1$, and $\delta=0.06$, the system locates in the SDW-M phase, the corresponding map of spectral weight is shown in Fig.~\ref{ferm1}(a). As doping
increases to $\delta=0.14$, the system locates in the PM-M phase, the corresponding spectral weight is depicted in Fig.~\ref{ferm1}(b), which is quite different to that of Fig.~\ref{ferm1}(a).

At the end of this section, two remarks should be noted that the phase diagram on the plane of $t_2$-$\delta$ is similar to that on the plane of $J_2/J_1$-$\delta$ shown in Fig.~\ref{fig5}. For a finite hopping $t_3$, the area of SDW-Ins phase will shrink remarkably, and the detailed results are presented in the appendix.

\section {temperature dependence calculations}\label{temperature_effects}

\begin{figure}
\centering
       \includegraphics[width=8cm]{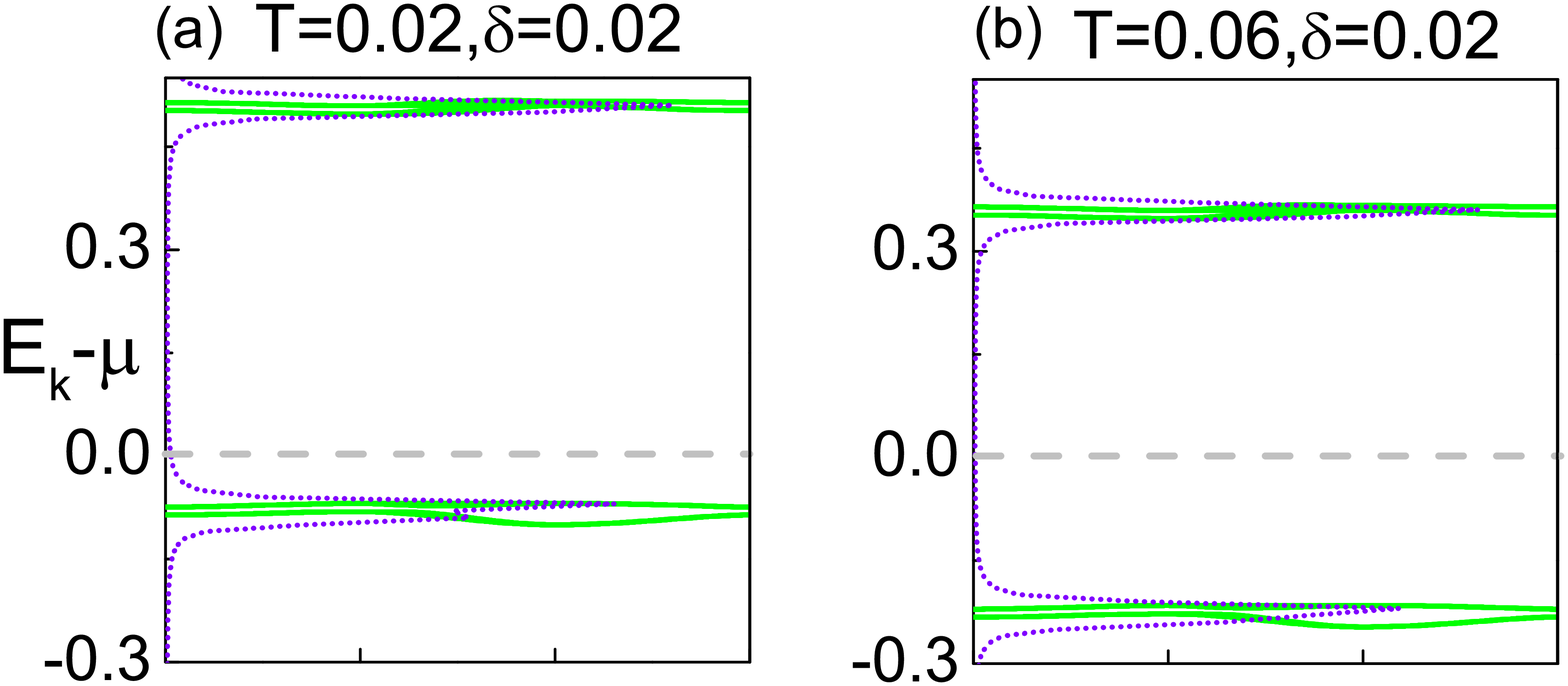}
       \includegraphics[width=8cm]{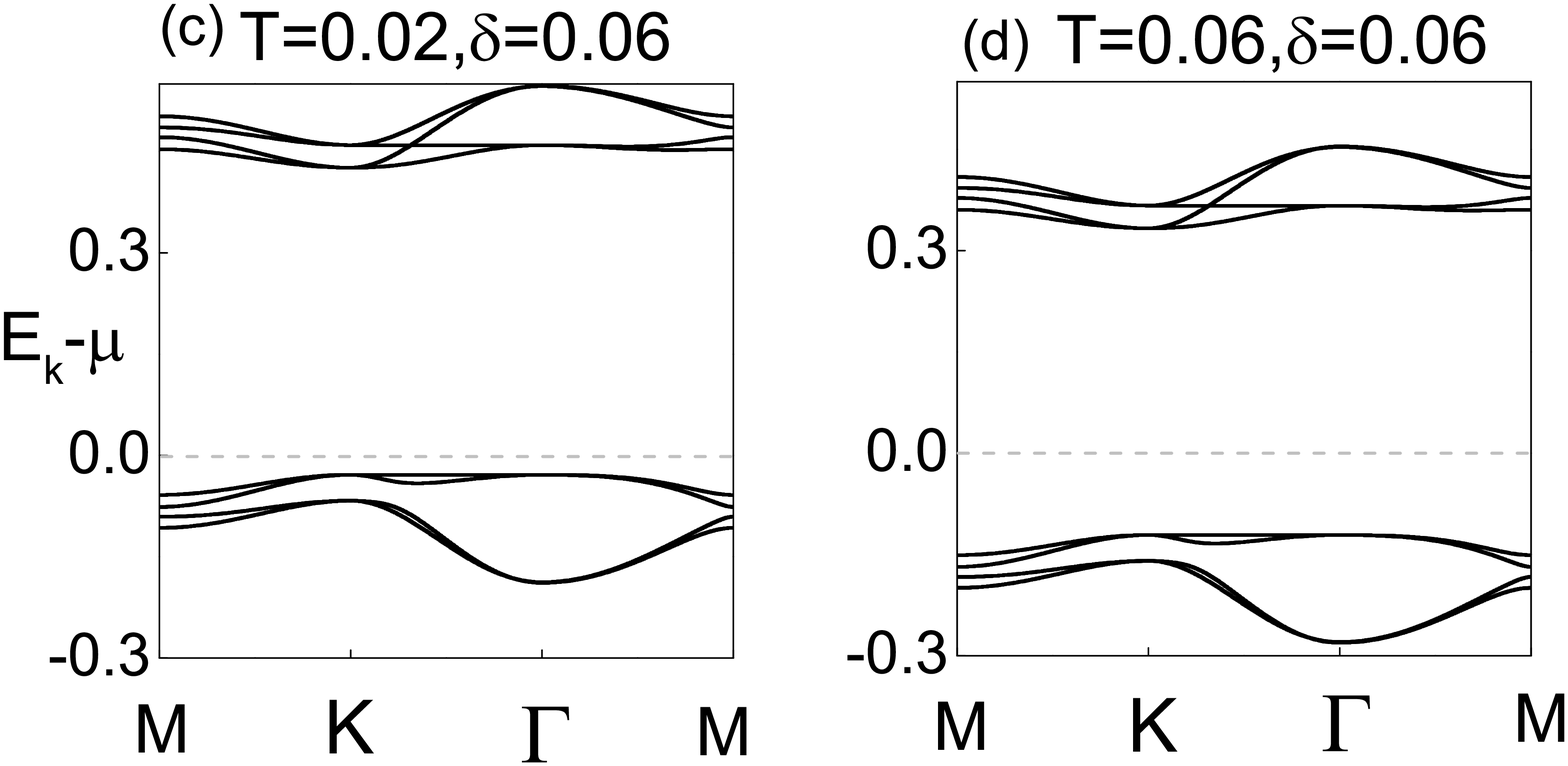}
\caption{(Color Online) For $t_2=0.3$ and $J_2=0$, the energy band dispersions along the high symmetric $\mathbf{k}$-point line M-K-$\Gamma$-M for different $T$ at $\delta=0.02$ and $\delta=0.06$, respectively. The violet dotted lines are the corresponding DOS for $\delta=0.02$, and the grey dashed lines denote the Fermi level.}\label{fig8}
\end{figure}

We then turn to discuss the temperature dependent properties using the parameters of $J_2=0$ and $t_2=0.3$. At sufficiently high temperature, $|m|$ will vanish, both the SDW-Ins and SDW-M phases will transit into the PM-M phase accompanying with the disappearance of $\Delta_{SDW}$ and $\Delta_{Fm}$.
We find that all bands move downward as $T$ is increased from $0.001$ to the critical temperature of magnetic order $T_c(m)$, accompanying with enlargement of $|\Delta_{Fm}|$. Seen from Fig.~\ref{fig8}, the lower subbands of SDW-Ins phase is deep-lying the Fermi level by increasing the temperature for $x=0.02$. For $\delta=0.06$, it is interesting to point out that the distinct Fermi surface of $T=0.001$ disappears ascribing to the increasing temperature. Besides the strong evidence from the presence of Fermi surface for metal, the electrical conductivity  $\sigma_{xx}=D\delta(\omega)$ for zero frequency electric field is also evaluated numerically to solidify the nature of a metal, where $\delta(\omega)$ is the delta function.

\begin{figure}
\centering
      \includegraphics[width=7.0cm]{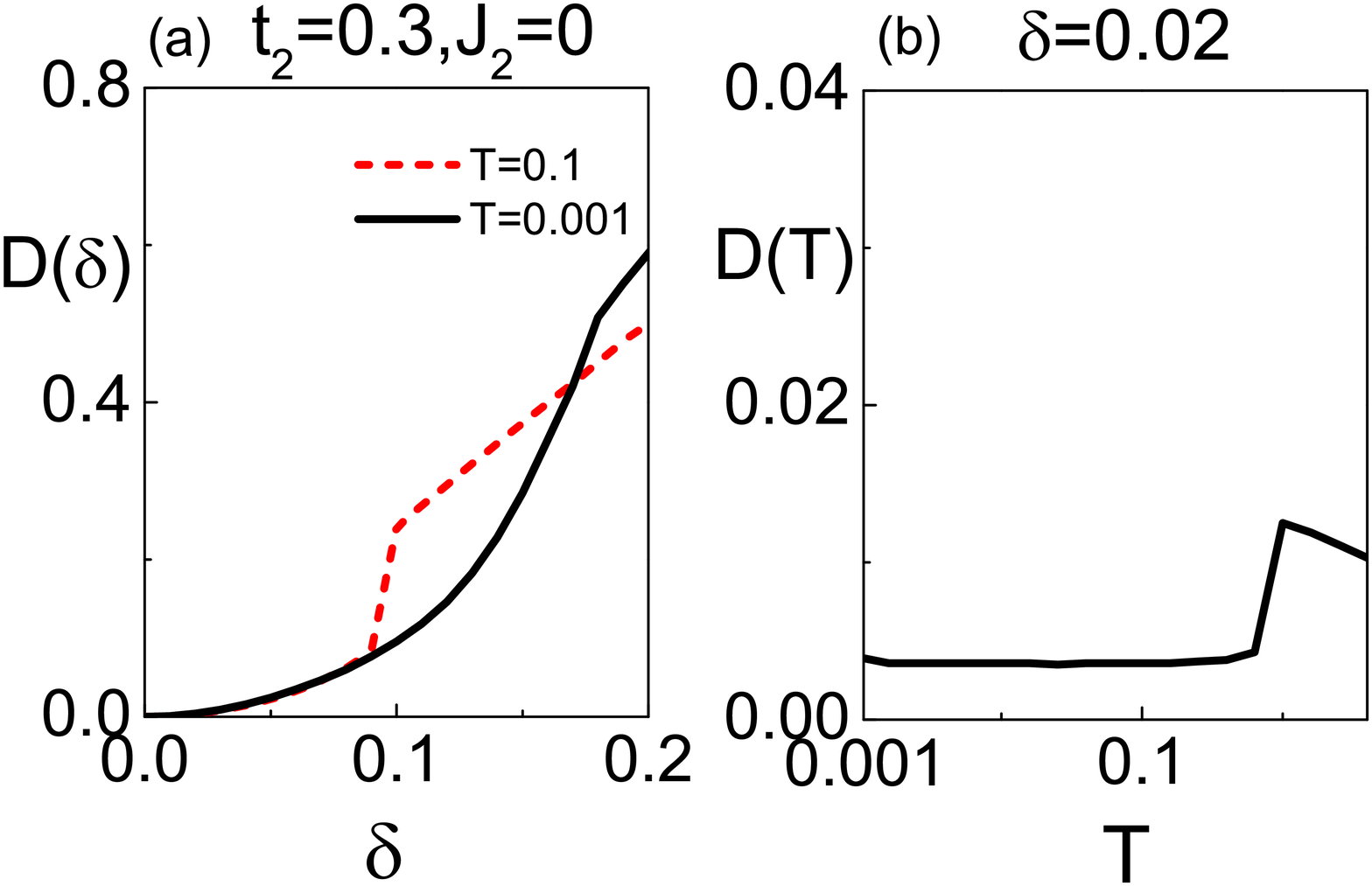}
\caption{(Color Online) For $t_2=0.3$ and $J_2=0$, (a) the Drude weight $D(\delta)$ as function of $\delta$ for $T=0.001$ and $0.1$, respectively. (b) The Drude weight $D(T)$ as function of $T$ with fixed $\delta=0.02$.}\label{fig10}
\end{figure}

For a single band system, the diamagnetic term determines the Drude weight. While for a multi-band system, the interplay among the bands below and above the Fermi level will induce a finite $\Pi_{xx}$ even at zero $T$ and is named as a geometric contribution.~\cite{fla2}
Fig.~\ref{fig10}(a) depicts the $D(\delta)$ as function of doping for $T=0.001$ and $T=0.1$, respectively.
In both cases the $D(\delta)$ monotonously increases as a function of doping, the higher doping levels correspond to the larger $D(\delta)$ and the more mobile charges. The Drude weight increases insignificantly at weak dopings, for $\delta=0.02$ it is $0.004$, and for $\delta=0.04$ it is $0.016$. For considerable large doping levels, the $|m|$ vanishes along with zero paramagnetic current due to the touching of energy subspaces, and the $D$ has a linear behavior versus doping. It is worth pointing out that the Drude weight is independent on the temperature at weak dopings since the two curves of $T=0.001$ and $0.1$ overlap for a small value of $\delta$.

Fig.~\ref{fig10}(b) shows the $D(T)$ as function of $T$ for $\delta=0.02$. In the SDW-Ins phase, the mobile charge remains unchanged value at various \textbf{$T<0.15$}.
As $T$ increasing up to $0.15$, the $D(T)$ jumps discontinuously and corresponds to a PM-M phase. Except for the half-filling, the $D$ is not exactly zero, instead of by a tiny value in the SDW-Ins phase.

\begin{figure}
\centering
      \includegraphics[width=4.0cm]{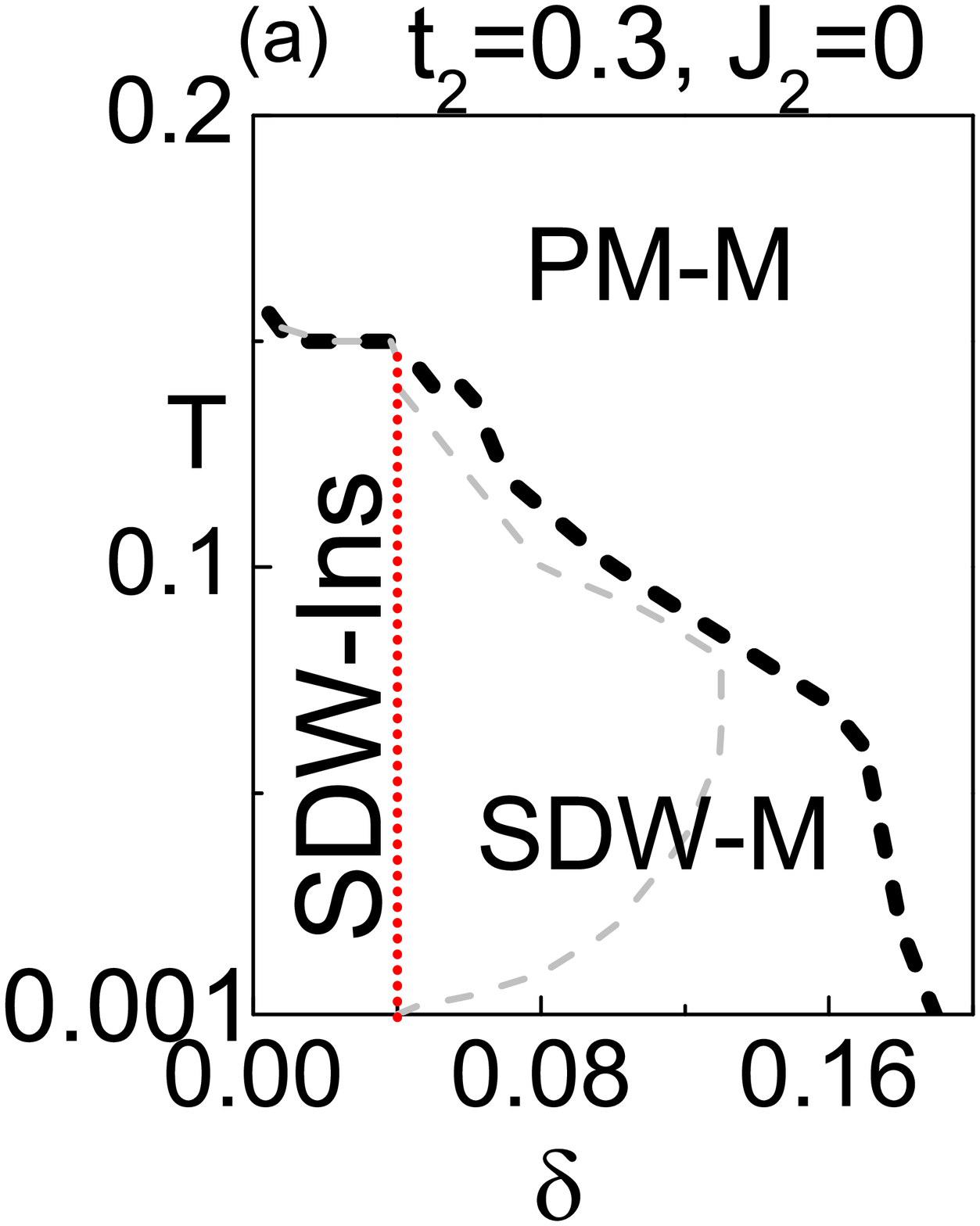}
       \includegraphics[width=4.0cm]{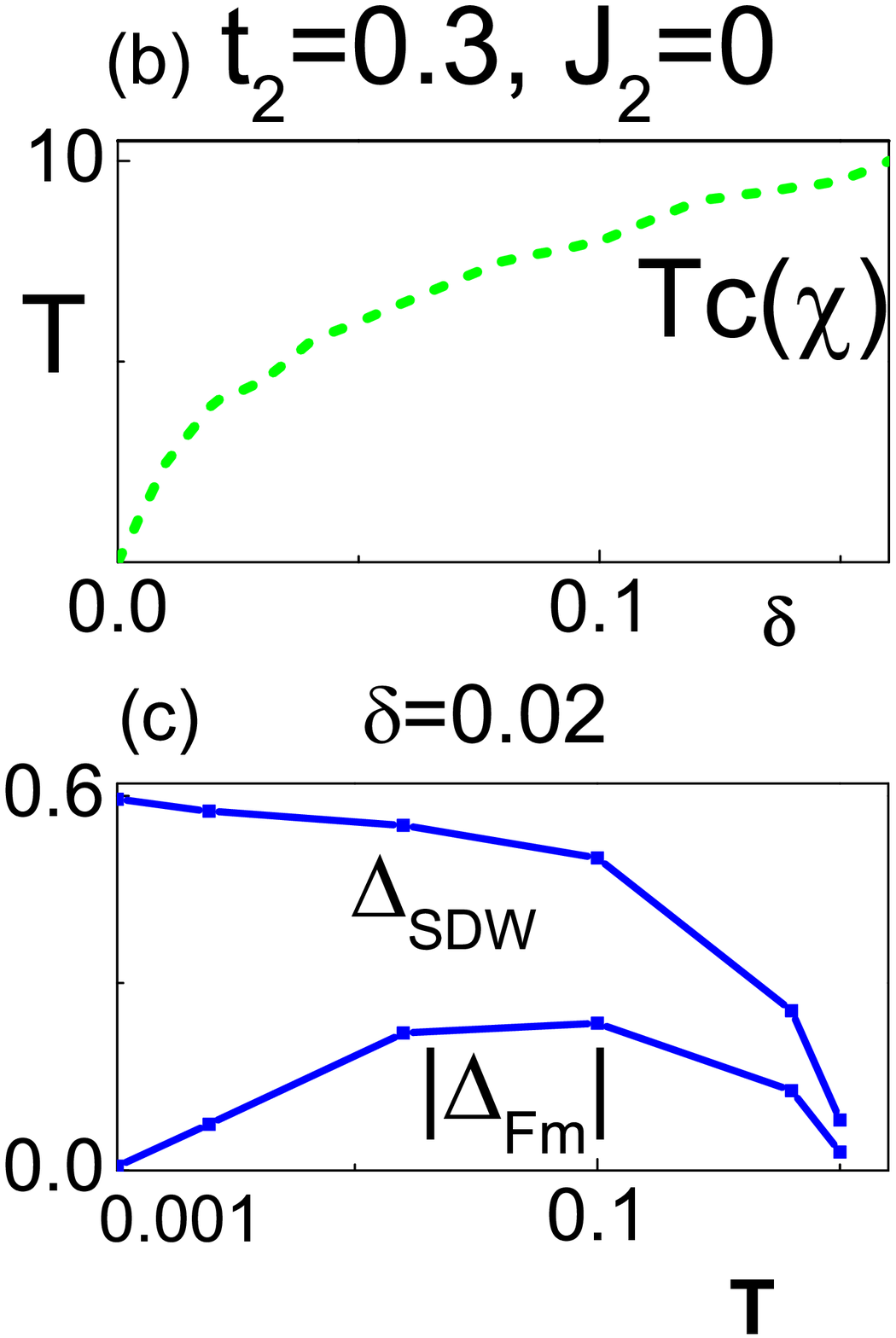}
\caption{(Color Online) (a) The phase diagram in $T$-$\delta$ plane for the parameters of $t_2=0.3$ and $J_2=0$. States in the left side of the grey dashed line denote the absence of Fermi surface. (b) Critical temperature of bond order $T_c(\chi)$ in $T$-$\delta$ plane. (c) $\Delta_{SDW}$ and $|\Delta_{Fm}|$
as a function of $T$ at the high symmetric point $M$ of the energy band $E_{4,k}$ for $\delta=0.02$.}\label{fig7}
\end{figure}

Phase diagram in the plane of $T$ vs $\delta$ is shown in Fig.~\ref{fig7}(a). As increasing the temperature $T$ from $0.001$, $|m|$ is obviously suppressed and approached zero at $T_c(m)$, whereas $\chi$ has negligible change in the range of [$0.001$,$T_c(m)$]. The system locates in the SDW-Ins phase for $x\leq0.04$ and $T<0.16$, with finite $\Delta_{SDW}$ and finite $\Delta_{Fm}$ as well as a small number of mobile charge. It will be transited into the PM-M phase when the temperature is beyond $T_c(m)$. If the doping level is larger than $0.04$, the system favors to stay in the SDW-M phase at low $T$, and then it will enter into the PM-M phase as the temperature is increased beyond $T_c(m)$, with large $\delta$ corresponding to low $T_c(m)$. In the SDW-M phase, the states located at the left side of the grey dashed line denote the absence of Fermi surface.

Fig.~\ref{fig7}(b) plots the critical temperature of $\chi$ denoted by $T_c(\chi)$ in the plane of T and $\delta$, above the curve all $\chi^{\sigma}_{ij}$ are zero, which will not influence on the division of the phase diagram in panel (a), since $T_c(\chi)$ is much higher than that of $m$. The movement of boson bond order $\langle b^{\dag}_{i}b_{j}\rangle$ follows that of fermion bond order $\chi^{\sigma}_{ij}$,~\cite{kp1,kp2,kp3} thus the Bosonic fluctuations only appear on a much higher energy scale and the Bonson condensation is assumed in our discussions.

Fig.~\ref{fig7}(c) depicts the $\Delta_{SDW}(M)$ and $|\Delta_{Fm}(M)|$ as function of $T$ at the high symmetric point $M$ on the flat band $E_{4,k}$ with fixed $\delta=0.02$. One can see that the $\Delta_{SDW}(M)$ monotonously decreases as $T$ increases,
while $|\Delta_{Fm}(M)|$ increases from $0.007$, and then drops as the system approaches the boundary of PM-M phase.

\section{summary}\label{summary}

SSL can be realized in a group of synthesized compound, where the exhibiting antiferromagnetic metallic phase has been reported within a specified range of parameters,~\cite{haidi} but the SDW-Ins phase defined in our discussions has rarely been mentioned since the flat-banded system has triggered intensive interests only recently by the realization of Lieb lattice. Based on SB mean-field theory we investigate the intriguing insulator-metal phase transition on SSL by using $t_1$-$t_2$-$J_1$-$J_2$ model. For the strongly correlated electron interactions, the lower bands and upper bands are well separated. SDW-Ins phase is resulted from the quasi-flat band in the range of $T=[0.001,0.15)$, $\delta\leq0.04$, displaying a finite $\Delta_{SDW}$ gap and a very tiny $D$ as well as the absence of Fermi surface. Only at half-filling, the Drude weight is exactly zero. The increasing of frustrated hopping $t_2$ and interaction $J_2$ almost are unaffected to the SDW-Ins phase.

The appearance of PM-M phase indicates the diminishing of magnetic order. The larger $t_2$ and $J_2$ as well as the larger doping will shrink the range of SDW-M phase since magnetic order is suppressed gradually. As $T$ is increased, the SDW-Ins phase and SDW-M phase will transit into the PM-M phase at the critical temperature. Although the Fermi surface of SDW-M phase will be immersed by higher $T$, due to the presence of the flat-band features, the Drude weight is robust against $T$.

Ascribing to the presence of the frustration and the special geometry of the lattice, SSL has a quasi-flat band, which localizes the electrons and leads to the appearance of SDW-Ins phase at low doping levels. The effects of doping, temperature, hopping and exchange coupling on the phase diagram of SSL are systemically studied, which provide a useful theoretical guidance for the understanding the nature of flat band and correlated insulating materials.

\section{acknowledgements}

We thank Prof. Yongping Zhang for discussions. This work was supported by the State Key Programs of China (Grant Nos. 2017YFA0304204 and 2016YFA0300504), the National Natural Science Foundation of China (Grant Nos. 51672171, 11625416, and 11774218), and the Natural Science Foundation from Jiangsu Province of China (Grant No. BK20160094). W.L. also acknowledges the start-up funding from Fudan University. The fund of the State Key Laboratory of Solidification Processing in NWPU (SKLSP201703), the supercomputing services from AM-HPC, and Fok Ying Tung education foundation are also acknowledged.


\begin{appendix}
\setcounter{figure}{0}
\renewcommand{\thefigure}{A\arabic{figure}}

\section{finite $t_3$ destroys SDW-Ins phase  }

\begin{figure}
\centering
      \includegraphics[width=4cm]{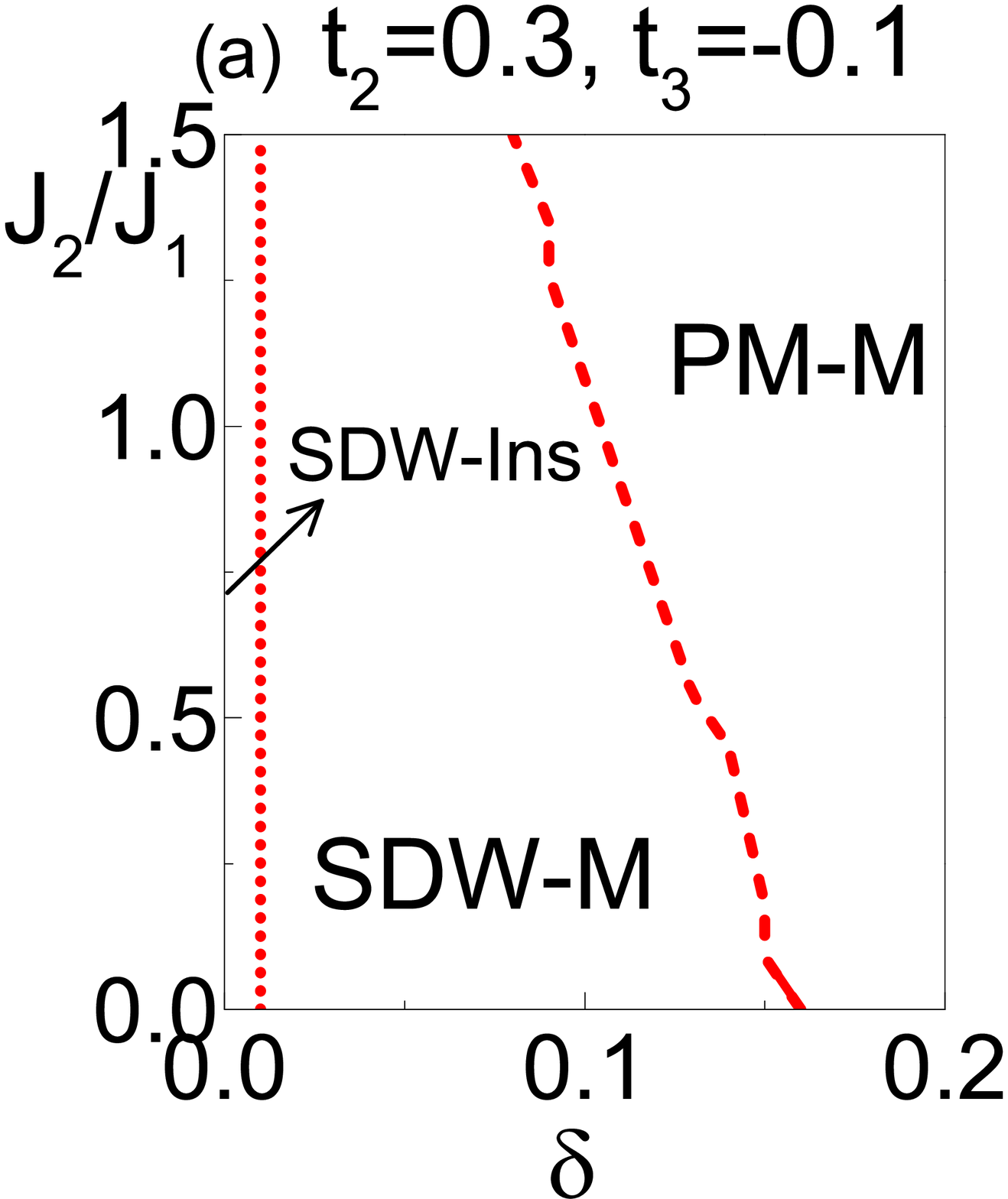}
      \includegraphics[width=4cm]{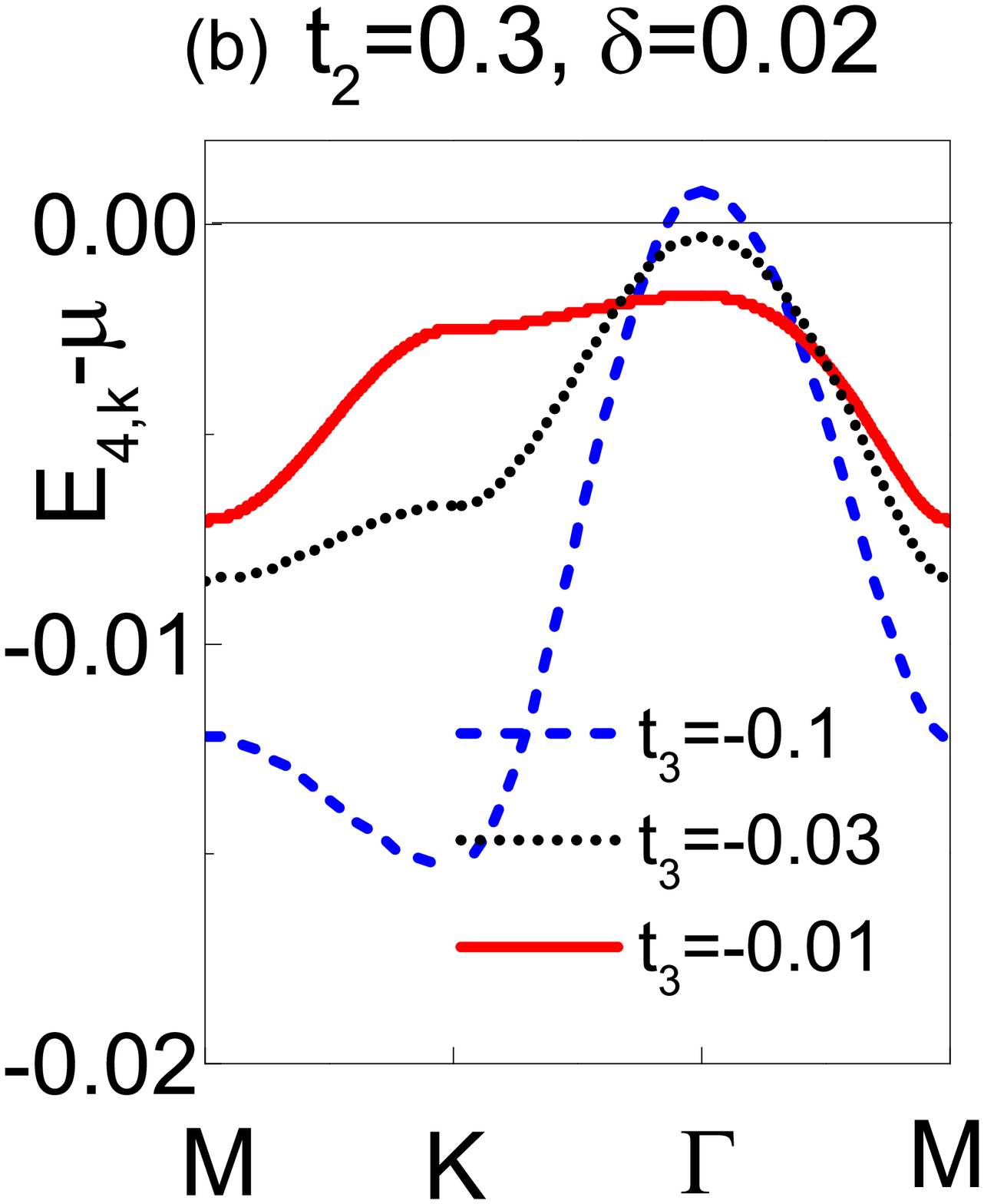}
\caption{(Color Online) For $T=0.001$ and $t_2=0.3$, (a) the phase diagram on $J_2/J_1$-$\delta$ plane for $t_3=-0.1$. (b) Variation of the top band of the low-energy space along the high symmetric $\mathbf{k}$-point line M-K-$\Gamma$-M for $J_2=0$ and $\delta=0.02$, where we take the different $t_3=-0.01,-0.03,-0.1$.}\label{fig11}
\end{figure}

\begin{figure}
      \includegraphics[width=9cm]{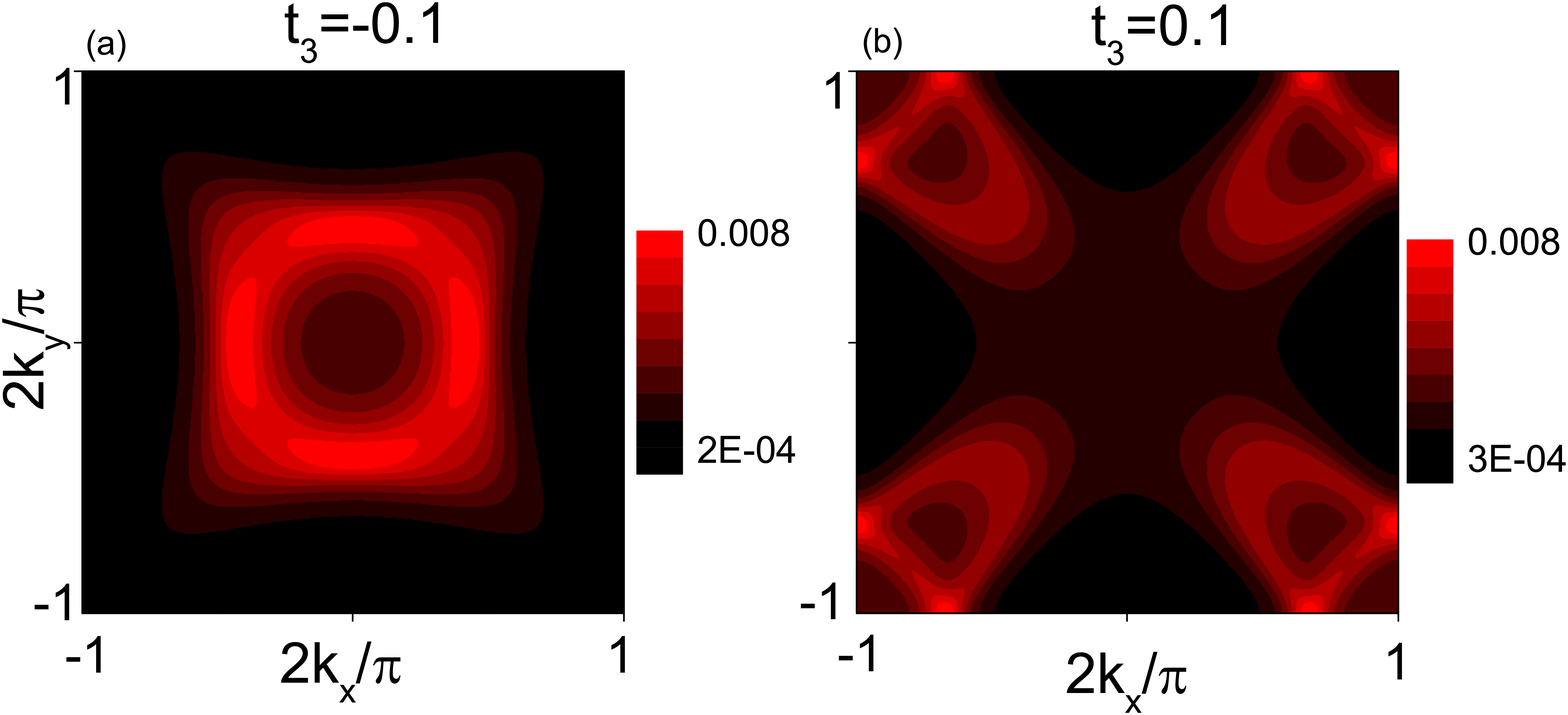}
\caption{(Color Online) The map of spectral weight of $\omega=0$ in the $\mathbf{k}$ space for $T=0.001, t_2=0.3, J_2=0.9J_1, \delta=0.06$ with (a) $t_3=-0.1$ and (b) $t_3=0.1$.}\label{fa3}
\end{figure}

Since a finite $t_3$ changes the curvature of the quasi-flat band, it will affect on the SDW-Ins phase dramatically. Here we find that SDW-Ins
phase almost vanishes as $t_3$ increased to $|t_3|=0.1$ at $T=0.001$, shown in Fig.~\ref{fig11}(a).
Fig.~\ref{fig11}(b) shows the topmost band in the lower subspace for various $t_3$ at $\delta=0.02$. At a small value of $t_3=-0.01$, the flat segment from K to $\Gamma$ is sloped, and the symmetry between M-K and $\Gamma$-M is broken. As further increasing $|t_3|$, the $E_{4,k}$ band will be dispersive significantly. Thus we conclude that a tiny value of $t_3$ will shrink the area of SDW-Ins phase remarkably due to the disturbing flat band.

Although both the positive and minus $t_3$ will change the curvature of the flat bands, they have a different topology of Fermi surface. For the minus value of $t_3=-0.1$, the pockets of Fermi surface locate at around the center point of $\Gamma$, while for the positive value of $t_3=0.1$, the Fermi surface becomes disconnected sheets at around the four points of $(\pm \pi/2,\pm \pi/2)$ in the first Brillouin zone.
The corresponding maps of the spectral weight also have quite different patterns, which are shown in Fig.~\ref{fa3}.

\section{electronic doping }

General speaking, SB theory is widely used in the hole-doped case where there is no double occupied site. In fact after particle-hole transformation, it can also be applied in electron-doped case.

Deviated from half-filling in electron-doped case, an electron can propagate from a double occupied site to a single occupied site $n_{i\bar{\sigma}}c^{\dag}_{i\sigma}c_{j\sigma}n_{j\bar{\sigma}}$, here $n_{j\bar{\sigma}}\approx \delta$ and $n_{i\bar{\sigma}}\approx 1$. After particle-hole transformation $c^{\dag}_{i\sigma}=c_{i\sigma}\xi_i$, the above expression can be rewritten as $-(1-n_{i\bar{\sigma}})c^{\dag}_{j\sigma}c_{i\sigma}(1-n_{j\bar{\sigma}})\xi_i\xi_j$, which looks like the form of hole-doped case. The sign of $\xi_i=\pm1$ depends on which sublattice $i$ belongs to. The four particle interaction Hamiltonian has the same form in the particle-hole transformation. Thus we can deal the electron- and hole- doped cases in the same manner as we choosing a special $\xi_i$.

On the mean-field described $t_1$-$t_2$-$J_1$-$J_2$ SSL, after the particle-hole transformation of $\xi_i=(-1)^i, i=1...4$, and taking $t_1=-t_1$ simultaneously,
kinetic Hamiltonian transforms as $H_{t,k}\rightarrow -H^*_{t,k}$. The interaction Hamiltonian can also be changed into $ H^{\sigma}_{J,k}\rightarrow-H^{*\sigma}_{J,k}$,
as long as mean-field $\chi$ changes as $\chi^{\sigma}_{i,j}\rightarrow\chi^{\sigma}_{i,j}$ on diagonal links, and $\chi^{\sigma}_{i,j}\rightarrow-\chi^{\sigma}_{i,j}$ on n.n. links.
In addition, it can be  proved that $H_k=H_k^*$ by changing $k_x\rightarrow-k_x,k_y\rightarrow-k_y$ in Eq.(\ref{1}).
Therefore particle-hole transformations along with minus sign of $t_1$ will transform $H_k\rightarrow -H_k$ in our discussions. Thus doping electrons actually means doping holes in our
treatment, with the energy band being inverted $E_n\rightarrow -E_n$.

\end{appendix}

\end{document}